\documentclass[%
 reprint,
 amsmath,amssymb,
 aps,
]{revtex4-2}

\usepackage{hyperref}
\usepackage{placeins}
\usepackage{graphicx}
\usepackage{dcolumn}
\usepackage{bm}

\begin{document}

\preprint{APS/123-QED}

\title{Two-particle cumulant distribution: a simulation study of higher moments\\}

\author{Satya Ranjan Nayak$^{1}$}
 \email{satyanayak@bhu.ac.in}

 \author{Akash Das$^{2}$}
 \email{24pnpo01@iiitdmj.ac.in}
\author{B. K. Singh$^{1,2}$}%
 \email{bksingh@bhu.ac.in}
 \email{director@iiitdmj.ac.in}
\affiliation{$^{1}$Department of Physics, Institute of Science,\\ Banaras Hindu University (BHU), Varanasi, 221005, INDIA. \\
$^{2}$Discipline of Natural Sciences, PDPM Indian Institute of Information Technology Design \& Manufacturing, Jabalpur-482005, India\\
}
\date{\today}

\date{\today}
\begin{abstract}

In this work, we have shown the two-particle correlations of charged hadrons in d-Au collisions at 200 GeV in PYTHIA8/Angantyr simulations. These correlations were studied at different multiplicities and pseudorapidity intervals. The two-particle correlations arise due to color reconnections, resonance decays, jet correlations, and hadronic rescattering. These correlations are inversely proportional to multiplicity but remain unaffected for larger pseudorapidity windows. We treated these correlations as distributions and calculated their skewness and kurtosis. The non-flow distributions deviate greatly from a Gaussian distribution and have high skewness and kurtosis. The ``true" elliptic flow distributions resemble Gaussian distributions; they have significantly lower skewness and kurtosis. We suggest that if the two-particle cumulant flow is treated as an event-by-event distribution, its skewness and kurtosis can be instrumental in distinguishing true flow and non-flow.

\end{abstract}

\maketitle


\section{\label{sec:level1}Introduction\protect\\ }

The ultra-relativistic heavy-ion collisions at BNL RHIC and CERN LHC form a deconfined state of partons called a quark-gluon plasma (QGP) \cite{Collins:1974ky,Cabibbo:1975ig,Chapline:1976gy}. The QGP is known to behave like a near-perfect fluid with a small shear viscosity \cite{Heinz:2013th}. It evolves hydrodynamically to give rise to one of its iconic signatures, the ``elliptic flow" \cite{Reisdorf:1997fx,STAR:2013ayu}. The elliptic flow refers to the transfer of spatial anisotropy of the overlap region into the momentum anisotropy of the final state hadrons. It is characterized by the elliptic flow coefficient $v_2$ \cite{Voloshin:1994mz,Snellings:2011sz,Poskanzer:1998yz,Huovinen:2001cy,Huovinen:2006jp},

\begin{equation}
    v_2=\cos 2(\phi -\psi)
\end{equation}
where $\phi$ is the azimuthal angle and $\psi$ is the reaction plane angle. Alternatively, the coefficient $v_2$ can be calculated using the two-particle cumulant \cite{STAR:2022pfn,Borghini:2000sa,Borghini:2001vi}, i.e.

\begin{equation}
    v^{2}_{2}\{2\}=\langle\langle\cos 2\Delta\phi\rangle\rangle
\end{equation}
where $\Delta\phi$ is the azimuthal angle difference between a particle pair, excluding the self-interaction terms, the $\langle\langle..\rangle\rangle$ refers to the average over all tracks and events. 

The two-particle cumulant flow is not necessarily related to the collision geometry \cite{Feng:2024eos,Borghini:2002mv}. Apart from the hydrodynamic flow, the correlation between mother and daughter particles, particles from a jet shower, and hadrons from a string fragmentation can give non-zero flow \cite{Jacobs:2004qv,Wang:2013qca,Andersson:1983jt}. Since the origin of such correlations is not hydrodynamic and does not relate to the collision geometry, they are usually referred to as ``non-flow". Although these correlations are short-ranged, they can significantly affect the elliptic flow measurements in low multiplicity events.

In the recent runs, RHIC collided gold ions with proton and deuteron beams in search of QGP droplets in the smaller collision systems \cite{PHENIX:2018lia,Nagle:2021rep,STAR:2015kak,PHENIX:2022nht,PHENIX:2021ubk}. The elliptic flow and triangular flow were observed in d-Au collisions by the PHENIX collaboration using two-particle cumulants. Unlike the Au-Au collisions, the low multiplicity of the d-Au collisions makes them highly susceptible to contamination by non-flow. Hence, a study dedicated to non-flow is required to better estimate ``true" elliptic flow in these collision systems. The major aim of this work is to study the non-flow correlations in different kinematic and multiplicity ranges, and check their implication on elliptic flow estimates.

We used the PYTHIA8/Angantyr \cite{Bierlich:2018xfw} model to study the non-flow correlations in d-Au collisions. The model simulates a nucleus-nucleus collision based on the interaction between the wounded nucleons without assuming the formation of the quark-gluon plasma. Since the Angantyr model does not have a QGP phase, it lacks true hydrodynamic flow. Hence, it is the best tool to study the non-flow correlations. Additionally, the model offers mechanisms such as color reconnections and hadronic rescattering, which are claimed to affect the two-particle cumulants.

This article is organized as follows: in Section II, we have provided a brief description of the Angantyr model, and in Section III, we have shown the multiplicity and pseudorapidity dependence of non-flow and the effect of different processes in non-flow estimates. We further discussed the skewness and kurtosis of the non-flow distribution. Finally, in Section IV, we have summarized our work.

\section{Angantyr}

PYTHIA is an all-purpose Monte Carlo event generator. The Angantyr extension now enables PYTHIA to simulate heavy-ion collisions \cite{Bierlich:2018xfw}. The Angantyr model was inspired by the FRITIOF program and the notion of wounded nucleons. Angantyr selects the individual subcollisions using a modified Glauber model, which incorporates the color fluctuations in the nucleon substructures, so-called ``Glauber-Gribov corrections" \cite{Alvioli:2013vk,Alvioli:2017wou}. The sub-collisions where the target and the projectile interact for the first time are marked as absorptive sub-collisions and treated similarly to a standard non-diffractive pp event. The subsequent interactions of the ``once wounded" nucleons are marked as wounded target/wounded projectile sub-collisions and treated like a modified single diffractive sub-collision. The modified single diffractive interaction refers to a non-diffractive collision between a proton and a pomeron. The double diffractive and elastic sub-collisions are treated accordingly, and all sub-collisions are combined to simulate a heavy ion event.

The softer interactions in PYTHIA are modelled using the Multi-Parton Interaction (MPI) framework \cite{Sjostrand:1987su}. After the Multi-Parton Interactions, the $q\bar{q}$ dipoles undergo a color reconnection to reduce the length of strings and control the multiplicity of the event. Unlike the default color reconnection, a spatially constrained color reconnection was recommended for Angantyr in Ref.~\cite{Lonnblad:2023stc} due to the limited range of strong interactions. The color reconnection is only allowed between the nearby $q\bar{q}$ dipoles ($<0.5 $ fm). Finally, the hadronization is performed using the LUND string fragmentation model.

\section{RESULTS AND DISCUSSIONS}
In this section, we have shown the PYTHIA8/Angantyr results for d-Au collisions at 200 GeV. The exact configurations of PYTHIA runs are specified in the text. In the results where the run configuration is not mentioned, the reader should assume the default setup. The quantity $c_2=\langle\cos 2\Delta\phi\rangle$ is denoted as non-flow in the results section. It is the two-particle cumulant of an entire event. It should not be confused with the two-particle cumulant of the individual particles; when the average of all events is mentioned, it is denoted as $\langle c_2\rangle$.  The $|\eta|$ refers to the rapidity span at mid-rapidity, for instance $|\eta|<0.5$ refers to $-0.25<\eta<0.25$ and so on. 
\begin{figure}[h]
\includegraphics[width=.5\textwidth]{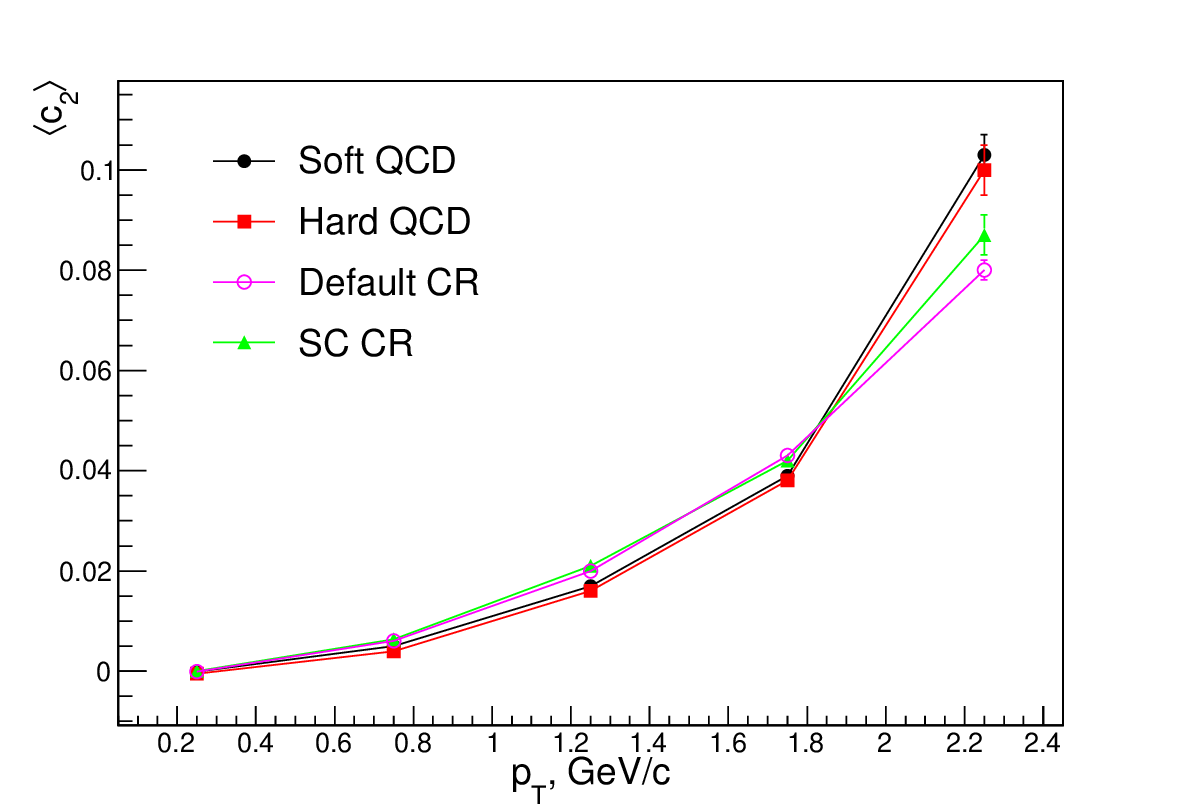}
\caption{\label{fig:3} The average two-particle cumulant $\langle c_2\rangle$ of charged hadrons with different configurations in d-Au collisions at 200 GeV. }
\end{figure}

\begin{figure}[h]
\includegraphics[width=.5\textwidth]{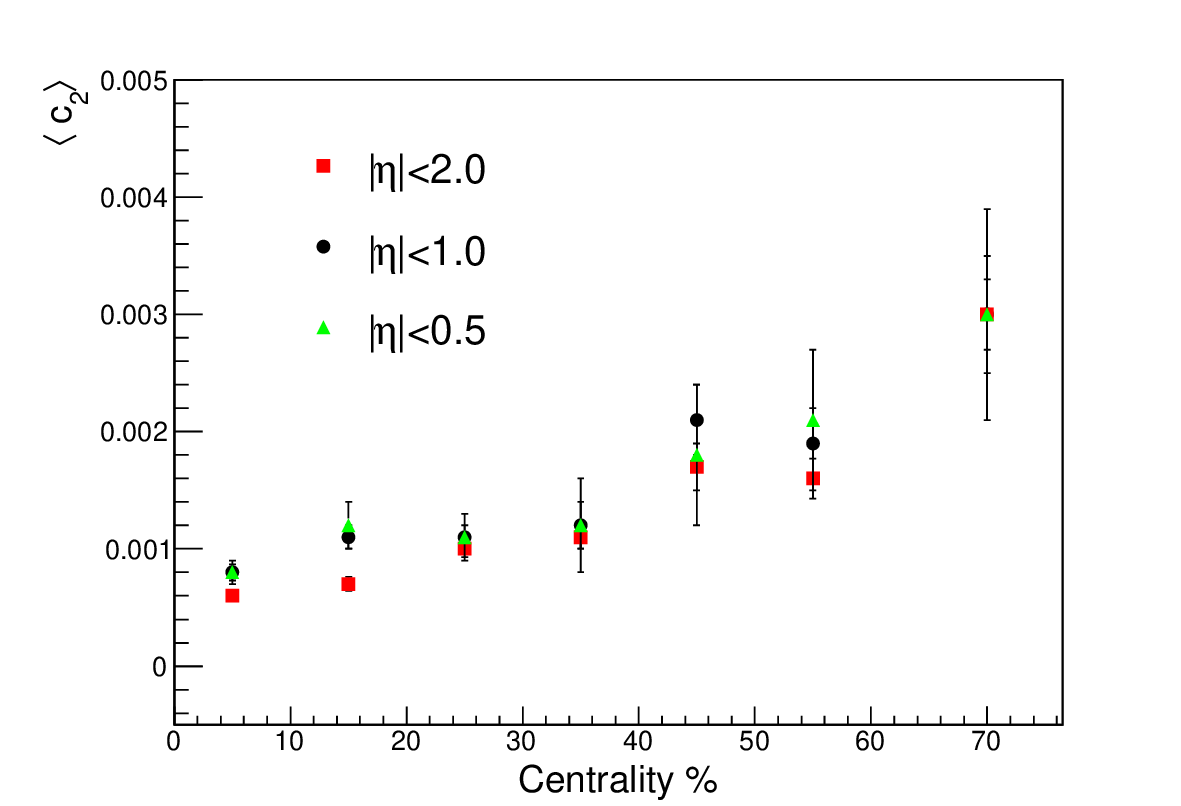}
\caption{\label{fig:3} The $\langle c_2\rangle$ of charged hadrons as a function of centrality in d-Au collisions at 200 GeV. }
\end{figure}

\begin{figure}[h]
\includegraphics[width=.5\textwidth]{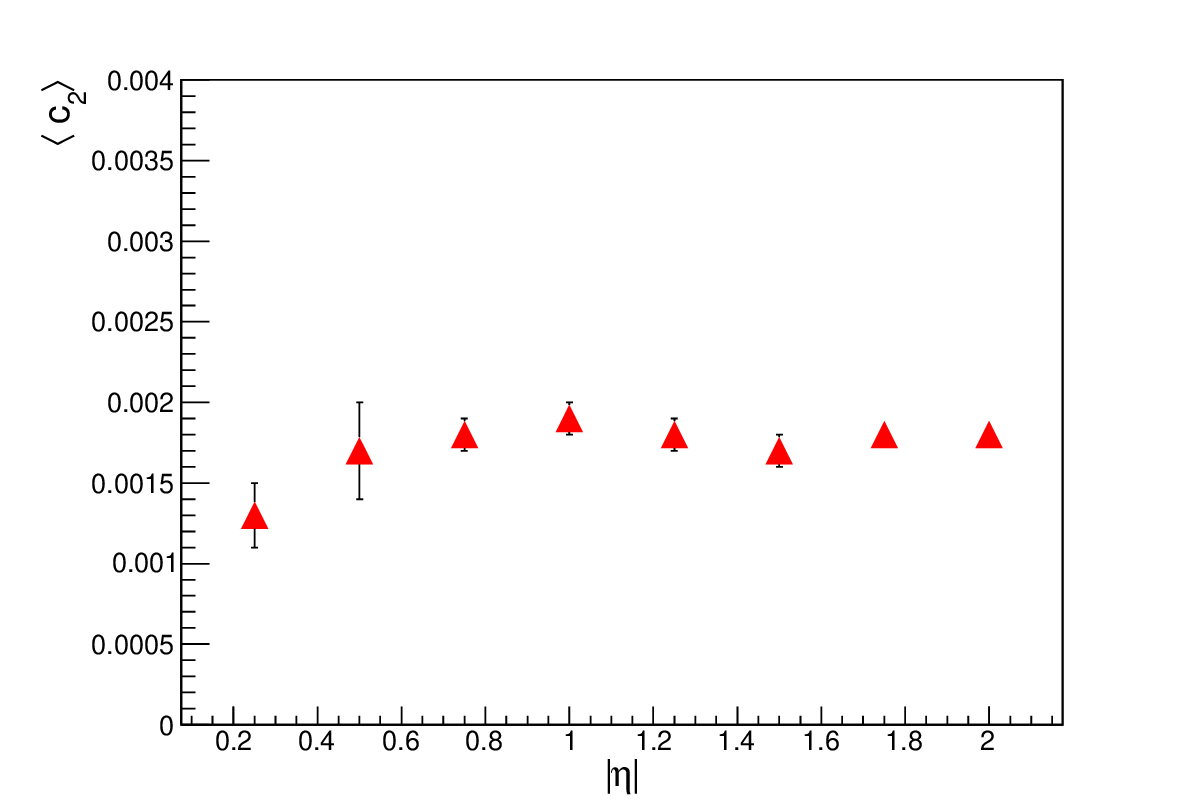}
\caption{\label{fig:3} The average two-particle cumulant $\langle c_2\rangle$ of charged hadrons for different $\eta$ windows in d-Au collisions at 200 GeV. }
\end{figure}

\begin{figure}[h]
\includegraphics[width=.5\textwidth]{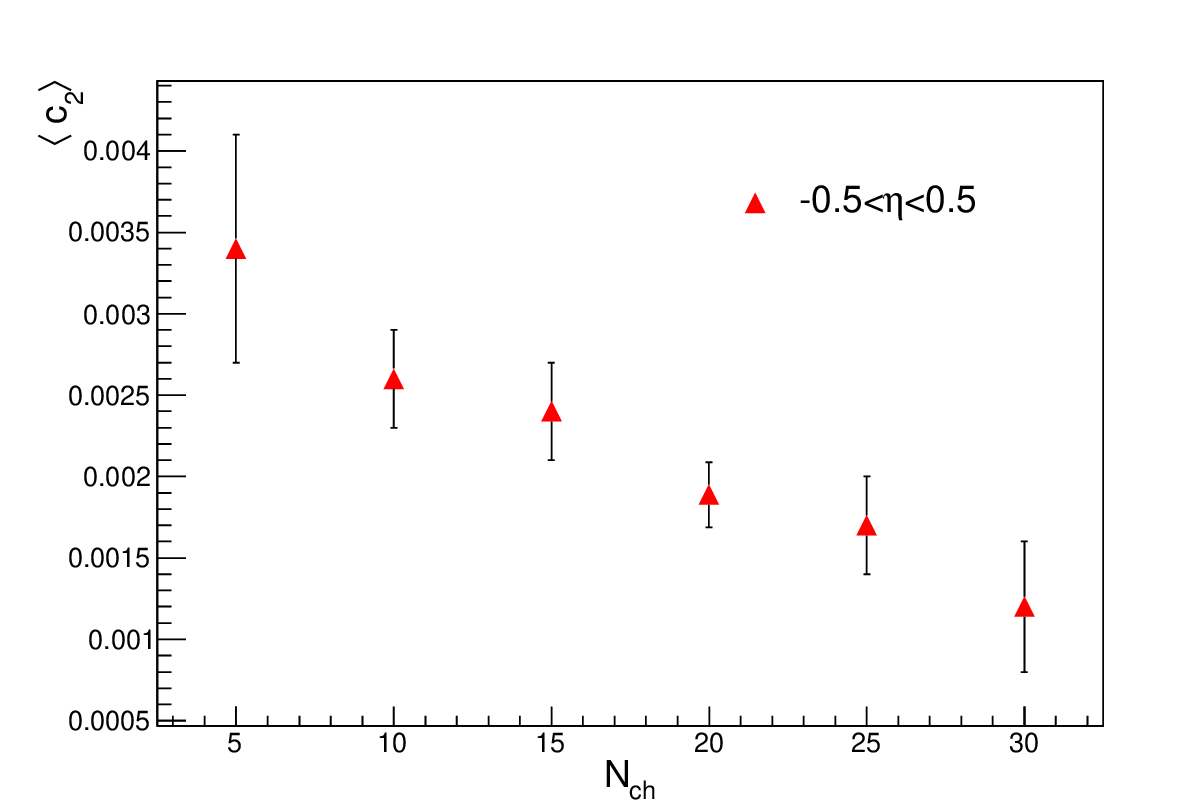}
\caption{\label{fig:3} The $\langle c_2\rangle$ of charged hadrons as a function of multiplicity in d-Au collisions at 200 GeV. }
\end{figure}

\begin{figure}[h]
\includegraphics[width=.5\textwidth]{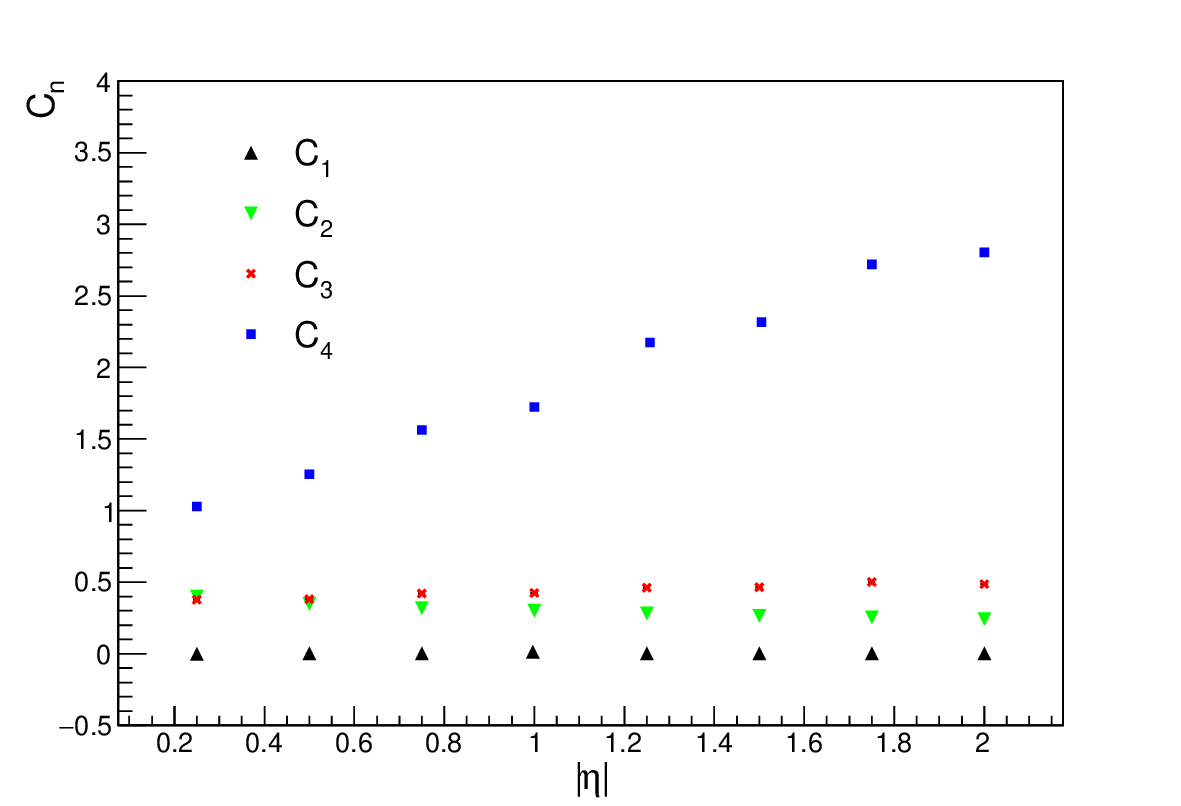}
\caption{\label{fig:3} The cumulants of $c_2$ distribution of charged hadrons in d-Au collisions at 200 GeV. }
\end{figure}

\begin{figure}[h]
\includegraphics[width=.5\textwidth]{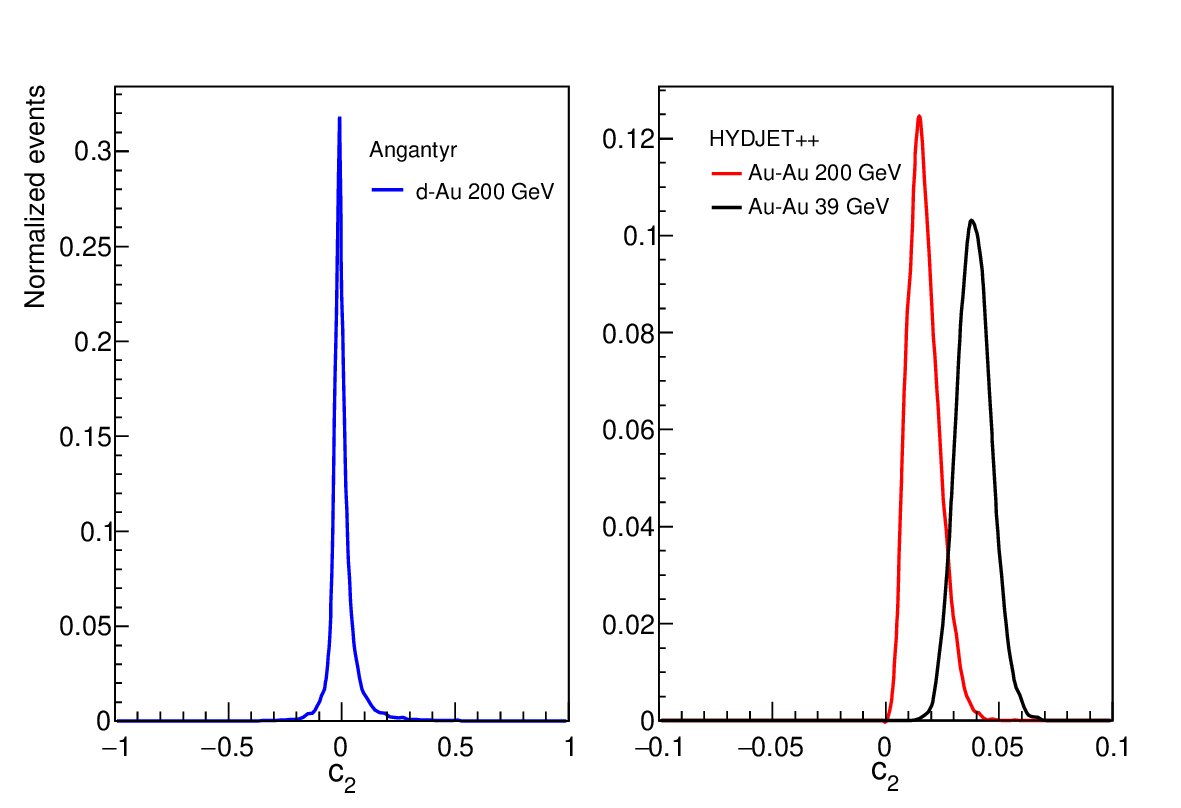}
\caption{\label{fig:3} The two-particle cumulant $c_2$ distribution of charged hadrons in d-Au collisions at 200 GeV, Au-Au collisions at 200 GeV, and Au-Au collisions at 39 GeV. }
\end{figure}

\begin{figure}[h]
\includegraphics[width=.5\textwidth]{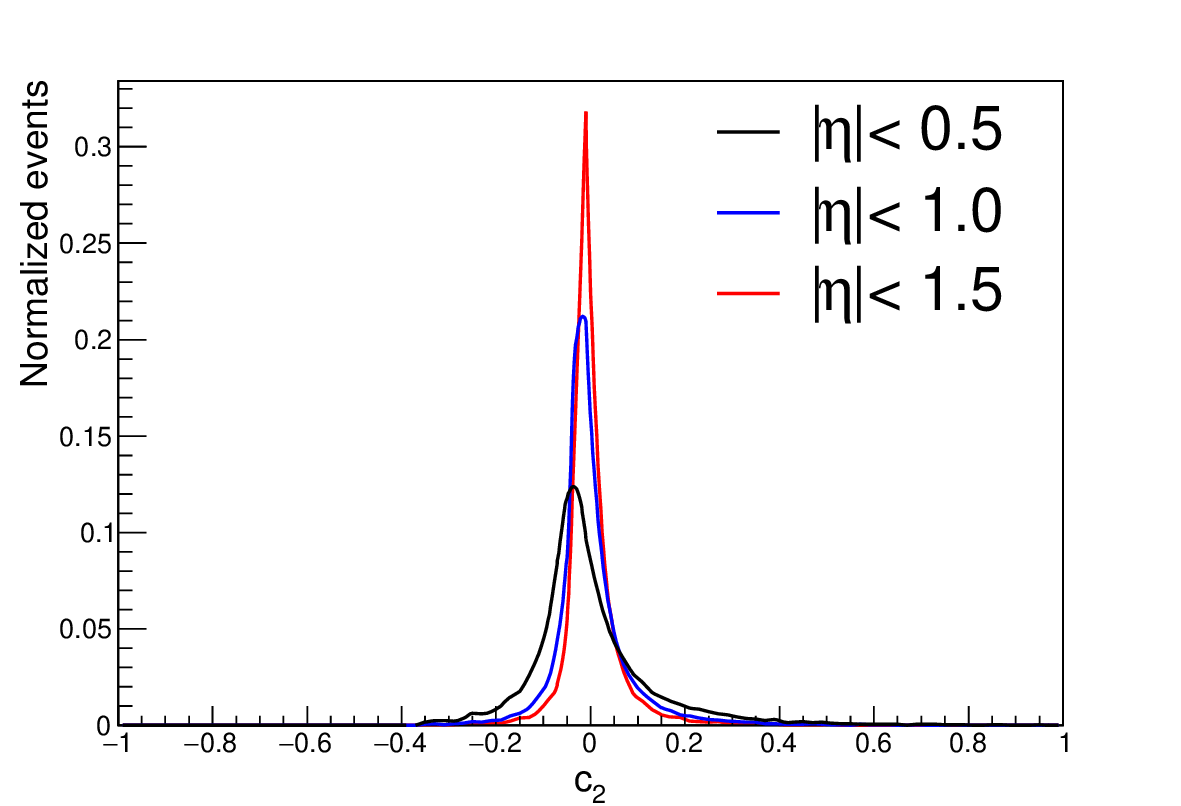}
\caption{\label{fig:3} The two-particle cumulant $c_2$ distribution of charged hadrons for different $\eta$ windows in d-Au collisions at 200 GeV. }
\end{figure}

\begin{figure}[h]
\includegraphics[width=.5\textwidth]{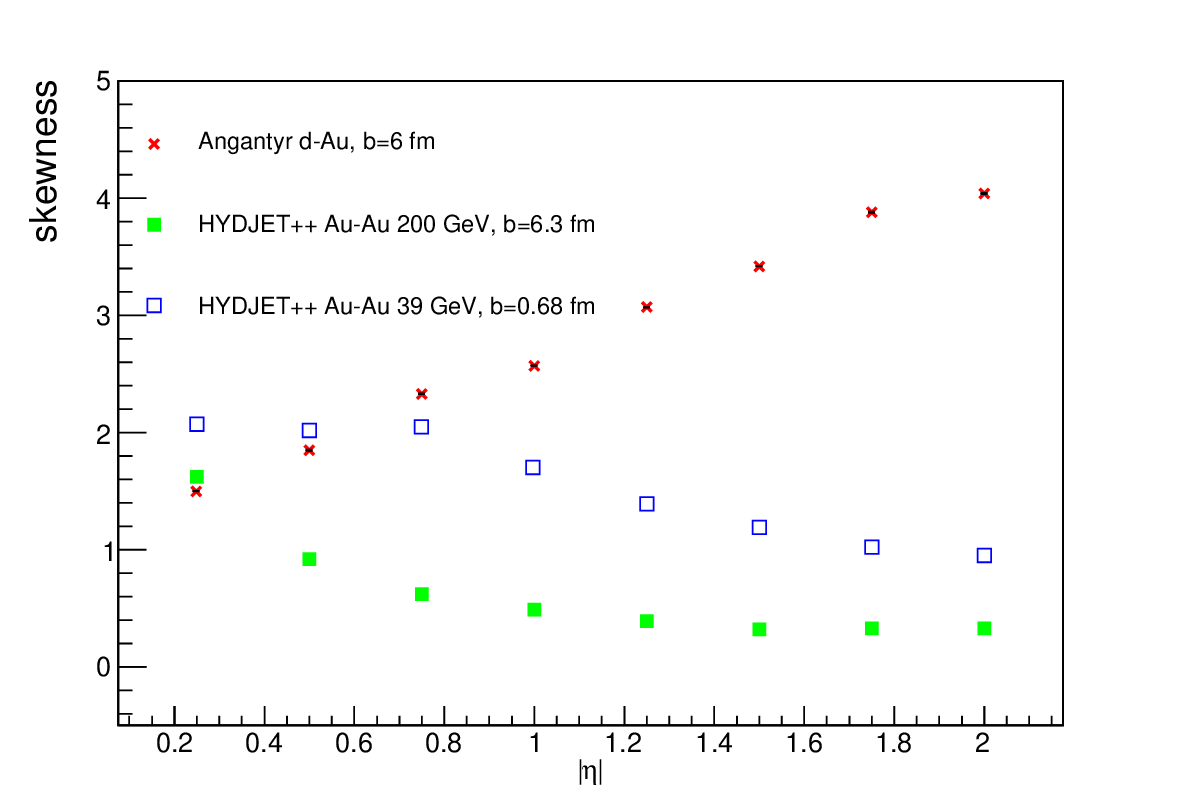}
\caption{\label{fig:3}  The skewness of non-flow distributions as a function of $|\eta|$ in d-Au collision at 200 GeV.}
\end{figure}

\begin{figure}[h]
\includegraphics[width=.5\textwidth]{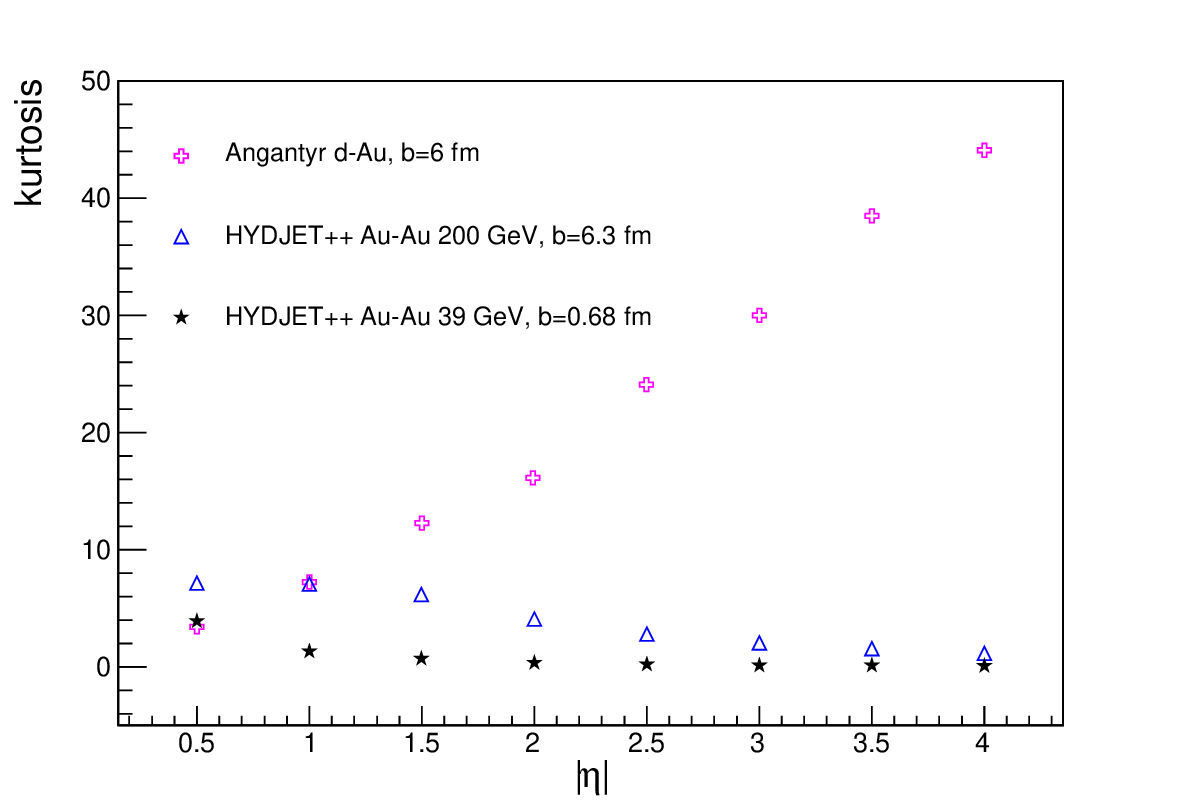}
\caption{\label{fig:3} The kurtosis of non-flow distributions as a function of $|\eta|$ d-Au in collision at 200 GeV. }
\end{figure}

\begin{figure}[h]
\includegraphics[width=.5\textwidth]{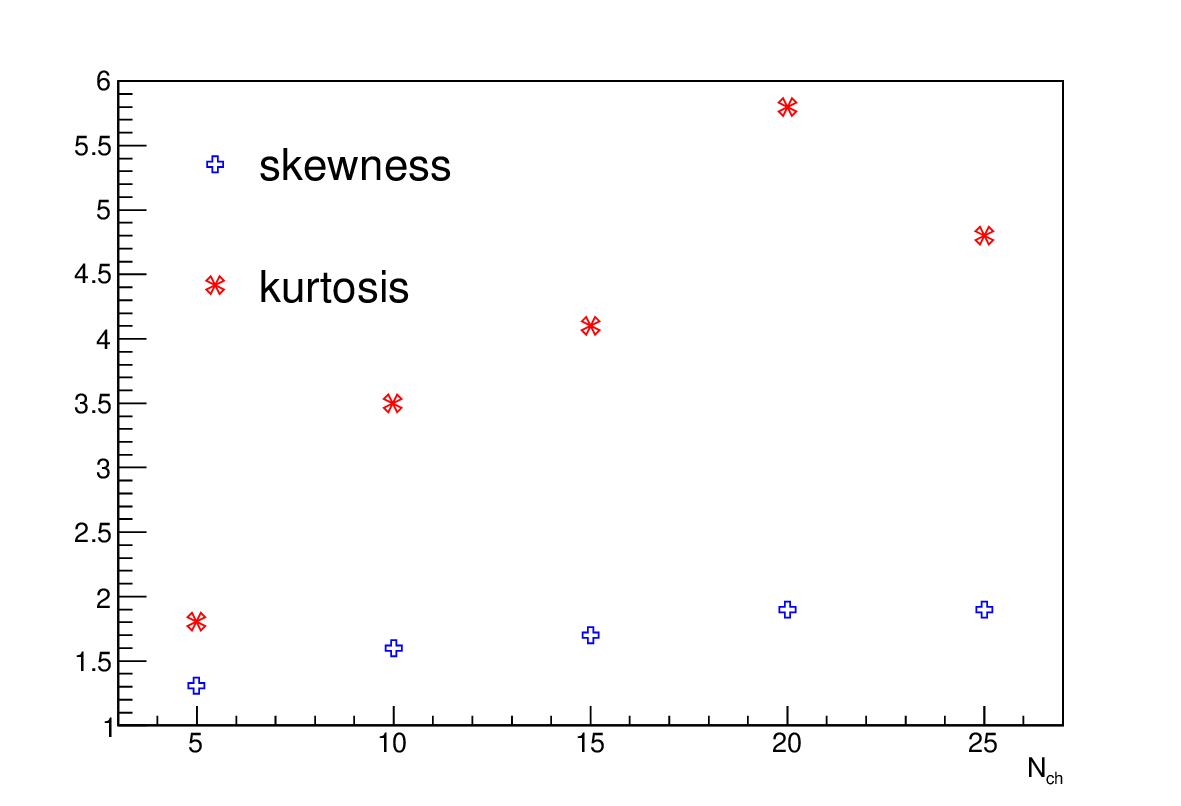}
\caption{\label{fig:3} The skewness and kurtosis of non-flow distributions as a function of multiplicity in d-Au collision at 200 GeV. }
\end{figure}
\subsection{Origin of two-particle correlations}

In PYTHIA/Angantyr simulations, the two-particle correlations emerge from softer interactions like MPIs, string fragmentation, color reconnections, and resonance decays. In Fig.~1, we have shown the average two-particle cumulant $\langle c_2\rangle$ of charged hadrons as a function of $p_T$ in minimum-bias d-Au collisions at 200 GeV with ``HardQCD:all=on" case, ``SoftQCD:all=on" case, ``Spatially constrained CR" case, and ``Hadronic Rescattering" case. The magnitude of $\langle c_2\rangle$ rises steadily with increasing $p_T$ in all configurations. This is expected, since most of the high-$p_T$ particles are produced by jets, which are one of the major contributors to non-flow. The simulations without the soft processes like MPI (HardQCD:all=on) have slightly smaller $\langle c_2\rangle$ values than ``SoftQCD:all=on" simulations, which suggests that Multi-Parton interactions lightly enhance the non-flow correlations.  The decays such as $\rho \rightarrow \pi^+\pi^-$, contribute significantly to the non-flow correlations. In Angantyr simulations, almost 15\% of the $\langle c_2\rangle$ originates due to resonance decays. A comprehensive study on the role of resonance decays in the two-particle correlations can be found in Ref.~\cite{Wang:2008gp}. The color reconnections mechanism is known to enhance two-particle correlations mimicking collective flow. The dataset with color reconnections has a higher $\langle c_2\rangle$ compared to the ones without color reconnections. Both the default color reconnections and the spatially constrained version of CR showed similar values of $\langle c_2\rangle$ at different $p_T$ ranges; a clear depiction can be found in Fig.~1. The color reconnections increase the multiplicity of high-$p_T$ particles and reduce $\langle c_2\rangle$ at high-$p_T$. The hadronic rescattering after the hadronization phase significantly alters the two-particle correlations of charged hadrons. The elastic scatterings pair-wise short-range momentum correlations, which increase the non-flow. The hadronic rescatterings produce a large number of short-lived resonances. These resonances decay shortly and contribute to non-flow. Hence, the ``Hadronic rescattering" configuration has the highest value of non-flow ($\langle c_2\rangle$).

In Fig.~2, we have shown the $\langle c_2\rangle$ of charged hadrons in d-Au collisions for different centrality bins. The coefficient $\langle c_2\rangle$ is lowest in the central collisions and increases rapidly in peripheral collisions. The centrality of each event was assigned based on the charged particle multiplicity in $-0.5<\eta<0.5$, similar to ref~\cite{Nayak:2025wco}. In the central collisions, the datasets with higher $|\eta|$ had a lower $\langle c_2 \rangle$. This can be attributed to the higher number of pairs with $\eta$-gaps. The $\langle c_2\rangle$ decreases with $|\eta|$, i.e, $\langle c_2\rangle \propto 1/N_{ch}$. The $\langle c_2\rangle$ as a function of $N_{ch}$ in $-0.5<\eta<0.5$ pseudorapidity range can be found in Fig.~4. However, in the minimum-bias collisions, the value of $\langle c_2\rangle$ is almost independent of $\eta$ windows. A clear representation of $\langle c_2\rangle$ as a function of $|\eta|$ in minimum-biased d-Au collisions can be found in Fig.~3. This suggests that larger $\eta$ windows do not guarantee non-flow removal. However, a $\eta$-gap can eliminate these correlations.

\subsection{Average value vs Distribution}
The two-particle cumulant ($c_2$) fluctuates for every event in Angantyr calculations. So, in this section,  we will treat $c_2$ as an event-by-event distribution rather than an event-averaged value. The true flow has a linear response to the initial spatial eccentricity. The event-by-event eccentricity distribution takes a Gaussian shape \cite{Voloshin:2007pc}. Ideally, the elliptic flow distribution should be a Gaussian \cite{Bhalerao:2018anl} with some deviations \cite{Yan:2014nsa}. The distribution of the average of random samples should be a Gaussian   \cite{ref1}. However, in the presence of non-flow correlations, events with higher $c_2$ are more likely, and the distribution develops a highly skewed tail. The low-multiplicity events are more likely to be affected, since a few jet correlations or decays can alter the overall $c_2$ of the event. These deviations can be instrumental in identifying ``true" hydrodynamic flow. The hydrodynamic flow and "non-flow" should show different trends in event-by-event fluctuations. The higher moments are a useful tool for studying event-by-event distributions. An effective method for examining event-by-event distributions is to use the higher moments. The $n^{th}$ moment of a distribution of $x$ is defined as,\\

\begin{equation}
    \mu_n =\langle (x-\mu)^n\rangle
\end{equation}

Where $\mu$ is the mean of the distribution. The cumulants of the distribution are defined as follows,\\

\begin{equation}
    C_1=\mu
\end{equation}

\begin{equation}
    C_2=\mu_2
\end{equation}

\begin{equation}
    C_3=\mu_3
\end{equation}

\begin{equation}
    C_4=\mu_4-3\mu_{3}^2
\end{equation}

The cumulants of the $c_2$ distribution in d-Au collisions as a function of $|\eta| $ window can be found in Fig.~5. The $2^{nd} $ cumulant (variance) decrease with $|\eta|$ and the higher cumulants like $C_3$ and $C_4$ increase with $|\eta|$. The standardized cumulants, like skewness and kurtosis, are dimensionless and scale-invariant, and provide a clear description of the shape of the distribution. Skewness refers to the skewed deviation of a distribution from a Gaussian distribution on either side of the mean. The kurtosis, on the other hand, suggests the ``peakedness" or the ``tailedness" of a distribution. The skewness and kurtosis are defined as follows,\\

\begin{equation}
    \textbf{skewness }(\gamma_1)=C_3/C_{2}^{3/2}
\end{equation}
\begin{equation}
    \textbf{kurtosis }(\gamma_2)=C_4/C_{2}^{2}
\end{equation}

The skewness arises from the $2^{nd}$ and $3^{rd}$ cumulants and kurtosis arises due to $2^{nd}$ and $4^{th} $ cumulants. The skewness and kurtosis also contribute to the higher-order cumulants like $C_6$ and $C_8$, i. e, $C_6=\mu_6-15(\gamma_2+3)\mu_{2}^3-10\gamma_1 ^2\mu_{2}^3+30\mu_{2}^3$ and $C_8=\mu_8-28\mu_2\mu_6-35(\gamma_2+3)^2\mu_{2}^4+420(\gamma_2+3)\mu_{2}^4+560\gamma_1 ^2\mu_{2}^4-630\mu_{2}^4$.

In order to compare with elliptic flow distributions, we have used the Monte-Carlo HYDJET++ calculations. The final state hadrons in HYDJET++ are located in a freeze-out hypersurface, which is modified using the spatial anisotropy parameter $\epsilon$. The flow rapidity profile of the final state particles is modified using the momentum anisotropy parameter $\delta$. The final state hadron distribution in HYDJET++ perfectly represents the distribution after a hydrodynamic expansion. A detailed description of the model and related topics can be found in the corresponding papers  \cite{Lokhtin:2008xi,Amelin:2006qe,Amelin:2007ic}. The parameters $\epsilon$ and $\delta$ were chosen as described in Ref.~\cite{Nayak:2024jbt}. It should be noted that the ``JET" component of the HYDJET++ contributes to non-flow correlations. However, the magnitude of $\langle c_2\rangle$ is significantly smaller than the overall contribution($\langle c_2 \rangle_{jets}\sim 10^{-5}$ and $\langle c_2 \rangle_{combined}\sim 10^{-3}$).  In Fig.~6, we have shown the event-by-event $c_2$ distribution of charged hadrons in d-Au collisions at 200 GeV using Angantyr for impact parameter $b=6 fm$. The $c_2$ distributions of charged hadrons in Au-Au collisions at 200 GeV ($b=6.3 fm$) and 39 GeV ($b=0.68fm$) obtained using HYDJET++ are shown for reference. The $c_2$ distribution from HYDJET++ is consistent with the Gaussian, while the $c_2$ distribution from Angantyr deviates from it greatly. These deviations are often overlooked while using event-averaged values of the respective observables. 


In Fig.~7, we have shown the $c_2$ distribution of charged hadrons at different $\eta$ windows. The distribution gets stiffer at higher $|\eta|$. Such dependence does not exist in hydrodynamic flow distributions. In Fig.~8, we have shown the skewness of the $c_2$ distribution from HYDJET++ along with Angantyr calculations. The skewness of the $c_2$ distribution is consistent with zero at higher $\eta$ windows, suggesting a smooth Gaussian distribution. The $c_2$ distribution from Angantyr is more right-shifted and has a higher positive skewness. The skewness of the $c_2$ distribution increases with $|\eta|$. This behavior is unique to non-flow distributions. It should be noted that the two-particle cumulant definition of elliptic flow should be used to check this aspect. The squared values of $\langle v_2\rangle$ calculated using the event plane method can not be used for such analysis. In doing so, the distribution will become highly skewed, almost like a half-Gaussian. The two-particle cumulant distribution in the presence of hydrodynamic flow only assumes a Gaussian shape in the case of a fixed impact parameter. A range of impact parameters will produce a distribution without a definite peak. The non-flow correlations do not depend on the impact parameter directly. So, the $c_2$ distribution from a non-hydrodynamic model will produce a skewed distribution both in fixed impact parameter collisions and minimum bias collisions. 

In Fig.~9, we have shown the kurtosis of the $c_2$ distribution in d-Au collisions at different $|\eta|$ windows. The kurtosis showed a linear increasing trend with $|\eta|$, similar to skewness. The $c_2$ distribution in Au-Au collisions from HYDJET++ decreased slightly at higher $|\eta|$s. This suggests that the ``true" elliptic flow distributions either remain unaffected by the change in $|\eta|$ windows, or show trends opposite to non-flow. A similar relation with $N_{ch} $ can be found in Fig.~10. The event-by-event fluctuations in flow distributions are often associated with initial state fluctuations. The characteristics of the ``skewness" and ``kurtosis" can be analyzed before such studies to check for non-flow contamination. A higher skewness/kurtosis, or a $|\eta|$ dependence, would suggest the presence of remnant non-flow. 

Many contributions have been made towards separating non-flow and elliptic flow in the past decade, including the use of higher-order cumulants and introducing a $\eta$-gap. Calculating higher-order cumulants requires heavy computing compared to calculating two-particle cumulants. Similarly, the use of $\eta$-gap method reduces the non-flow contamination. However, it has no scope for cross-checking residual non-flow after calculation. The statistical properties of the two-particle cumulant distribution can be used as a possible alternative since it is computationally light and provides scope for double-checking.

\section{Summary}
In the current letter, we have simulated d-Au collisions using PYTHIA8/Angantyr and analyzed the two-particle non-flow correlations. The two-particle cumulant $c_2$ emerges from different processes in Angantyr calculations. The major contributors are jet correlations and resonance decays. The softer processes under the MPI framework also contribute to the non-flow due to ``jet-like" correlations of low-energy partons. Such correlations have a higher $|\eta|$ range similar to ``true" flow. The Spatially Constrained Color Reconnection enhances these correlations, but its contribution is smaller than the default CR. The current implementation of hadronic rescattering significantly increases the two-particle correlations. The non-flow coefficient $c_2$ decreases rapidly with multiplicity/centrality. 

It is a common practice in heavy-ion physics to use the event-averaged values of observables like $v_2$. Alternatively, these observables can be treated as distributions. The ``hydrodynamic" flow distributions usually assume a Gaussian shape with some minor deviations. On the other hand, the non-flow distributions deviate significantly from the Gaussian distributions. These deviations can be characterized by the skewness and kurtosis of the respective distributions. We used the Au-Au collisions from HYDJET++ to represent ``true" flow distributions, and d-Au results from Angantyr were used for non-flow distributions. The skewness and kurtosis of non-flow distributions were significantly higher than those of the flow distributions from HYDJET++. Both skewness and kurtosis of non-flow distribution showed a linear relation with $N_{ch}$ and $|\eta|$ windows. This trend is unique to non-flow distributions. Hence, we conclude that a higher skewness/kurtosis or their $|\eta|$ dependence can be an indication of non-flow contamination.

\begin{acknowledgments}
BKS sincerely acknowledges financial support from the Institute of Eminence (IoE) BHU Grant number 6031. SRN acknowledges the financial support from the UGC Non-NET fellowship and IoE research incentive during the research work.  AD acknowledges an institute fellowship from PDPM Indian Institute of Information Technology Design \& Manufacturing, Jabalpur, India.

\end{acknowledgments}


\end{document}